\documentclass[10pt,conference]{IEEEtran}
\IEEEoverridecommandlockouts

\usepackage[numbers]{natbib}
\usepackage{amsmath,amssymb,amsfonts}
\usepackage{algorithmic}
\usepackage{graphicx}
\usepackage{textcomp}
\usepackage{multirow}
\usepackage{color}
\usepackage[table, dvipsnames]{xcolor}
\usepackage{makecell,cellspace,caption}
\usepackage[colorlinks,linkcolor=black,citecolor=black,urlcolor=black]{hyperref}
\usepackage{bbding}
\usepackage{fontawesome,numprint}
\usepackage{algorithm}
\usepackage{algorithmic}
\usepackage{amsmath}
\usepackage{pifont}
\usepackage{amssymb}
\usepackage{siunitx}
\usepackage[inline]{enumitem}
\usepackage{todonotes}
\setuptodonotes{inline}
\usepackage{mdframed}
\usepackage{subcaption}
\usepackage{float}
\usepackage{newfloat,caption,subcaption,listings,mdframed}
\usepackage{xspace}
\usepackage{balance}

\def\BibTeX{{\rm B\kern-.05em{\sc i\kern-.025em b}\kern-.08em
    T\kern-.1667em\lower.7ex\hbox{E}\kern-.125emX}}

\newcommand{\xmark}{\ding{55}}%
\newcommand{\approach}{PreciseBugCollector\xspace}%
\usepackage{xcolor,colortbl}

\definecolor{Gray}{gray}{0.85}
\definecolor{LightCyan}{rgb}{0.88,1,1}

\newcolumntype{a}{>{\columncolor{Gray}}r}
\newcolumntype{b}{>{\columncolor{white}}r}

\definecolor{bluekeywords}{rgb}{0.13,0.13,1}
\definecolor{greencomments}{rgb}{0,0.55,0.2}
\definecolor{redstrings}{rgb}{0.9,0,0}

\DeclareFloatingEnvironment[fileext=frm,placement={!ht},name=Listing]{listing}
\usepackage{fourier} 
\usepackage{array}
\usepackage{makecell}

\begin{document}

\pagestyle{plain}

\title{PreciseBugCollector:  Extensible, Executable and Precise Bug-fix Collection \\

}

\author{\IEEEauthorblockN{He Ye}
\IEEEauthorblockA{\textit{Carnegie Mellon University}\\
Pittsburgh, US \\
hey@cs.cmu.edu}
\and
\IEEEauthorblockN{Zimin Chen}
\IEEEauthorblockA{\textit{KTH Royal Institute of Technology}\\
Stockholm, Sweden \\
zimin@kth.se}
\and
\IEEEauthorblockN{Claire Le Goues}
\IEEEauthorblockA{\textit{Carnegie Mellon University}\\
Pittsburgh, US \\
clegoues@cs.cmu.edu}
}

\maketitle

\begin{abstract}
Bug datasets are vital for enabling deep learning techniques to address software maintenance tasks related to bugs. However, existing bug datasets suffer from precise and scale limitations: they are either small-scale but precise with manual validation or large-scale but imprecise with simple commit message processing.
In this paper, we introduce \approach, a precise, multi-language bug collection approach that overcomes these two limitations. \approach is based on two novel components: a) A bug tracker to map the codebase repositories with external bug repositories to trace bug type information, and b) A bug injector to generate project-specific bugs by injecting noise into the correct codebases and then executing them against their test suites to obtain test failure messages.

We implement \approach against three sources: 1) A bug tracker that links to the national vulnerability data set (NVD) to collect general-wise vulnerabilities, 2) A bug tracker that links to OSS-Fuzz to collect general-wise bugs, and 3) A bug injector based on 16 injection rules to generate project-wise bugs.
To date, \approach comprises \numprint{1057818} bugs extracted from \numprint{2968} open-source projects. Of these, \numprint{12602} bugs are sourced from bug repositories (NVD and OSS-Fuzz), while the remaining \numprint{1045216} project-specific bugs are generated by the bug injector.
Considering the challenge objectives,  we argue that a bug injection approach is highly valuable for the industrial setting, since project-specific bugs align with domain knowledge, share the same codebase, and adhere to the coding style employed in industrial projects.
\end{abstract}


\begin{table*}[t!]
\renewcommand{\arraystretch}{1.8}
\small
\caption{Bug Dataset in the Literature.}

\resizebox{\textwidth}{!}{\begin{tabular}
{llcccccrccccc}
\hline
 Bug Dataset & Publish Venue  & Publish Year &   Languages & Type &   Tests & Source &\# Bugs\\
\hline

Defects4J \cite{defects4j} &   ISSTA & 2014 & Java &  \checkmark &   \checkmark &commits+text processing & 835        \\
 ManyBugs \cite{LeGoues15tse} & TSE & 2015 & C & \checkmark& \checkmark & commits+text processing &   185 \\

IntroClass \cite{LeGoues15tse} & TSE & 2015 & C & \checkmark& \checkmark & student assignments &   998 \\
 
 CodeFlaw \cite{Codeflaws} &  ICSE-Companion & 2017 & C &  \checkmark & \checkmark &  contest     &  3902 & \\
 QuixBugs  \cite{quixbugs} &    SPLASH-Companion & 2017  & Java/Python & \checkmark & \checkmark & contest & 40    \\

CodRep \cite{Chen2018Coderep}  & Arxiv   &  2018   & Java    & \xmark  & \xmark  & commits+text processing & \numprint{58069}   \\

PatchPaser \cite{liu2018closer} & ICSME   &  2018   & Java    & \checkmark  & \xmark  & commits+text processing & \numprint{16450}   \\

 Bears \cite{Bears} & SANER & 2019 & Java & \checkmark & \checkmark &  commits+CI  & 251 \\
 BugsJS \cite{bugsjs-icst19}  & ICST & 2019 & JavaScript & \checkmark & \checkmark &  commits+text processing  &   453 \\
 Bugswarm \cite{bugswarm-icse19}  & ICSE & 2019 &   Java/Python &   \checkmark & \checkmark &  commits+CI &    \numprint{3091}  \\

 Defexts \cite{Defexts-icse19} & ICSE-Companion & 2019 &      Kotlin/Groovy & \checkmark & \checkmark&commits+text processing & 526 \\

 Refactory  \cite{refactory-ase19} & ASE & 2019  & Python & \checkmark & \checkmark & student assignments & 1783 \\

 BugsInPy \cite{BugsInPy-fse20} & FSE & 2020 & Python &    \checkmark & \checkmark   & commits+text processing  & 493 \\

ManySStuBs4J \cite{manystupidbugs} & MSR &  2020 & Java & \checkmark  & \xmark  & commits+text processing &  \numprint{153652} \\

CODIT's dataset \cite{codit-tse20} & TSE & 2020 & Java &  \xmark & \xmark   & commits+CI & \numprint{32473} \\

CodeBERT's dataset \cite{feng-etal-2020-codebert} & EMNLP & 2020 &Java/Python/Ruby/JavaScript/Go &  \xmark & \xmark &commits+text processing& 2 millions \\

CoCoNuT's dataset \cite{CoCoNuT} & ISSTA & 2020 &  Java/Python/C/JavaScript   &  \xmark & \xmark  &        commits+text processing &  23 millions \\

Megadiff \cite{megadiff} & Arxiv & 2021 & Java & \xmark  & \xmark  &   commits+text processing & \numprint{663029} \\

 Vul4J \cite{vul4j2022} & MSR & 2022 & Java &  \checkmark & \checkmark & commits+text processing  &    79 \\

FixJS \cite{fixjs-msr22} & MSR & 2022 & JavaScript  & \xmark  & \xmark  &  commits+text processing & 300 000 \\

\hline


\end{tabular}}

\label{background-dataset}
\end{table*}

\section{Introduction}
\label{sec:intro}

A precise bug dataset plays a crucial role in various software tasks, including bug detection  \cite{bug-detection-Khomh,bug-detection-2015}, 
fault localization \cite{jit-bug-detection-tse22,GZoltar,fl-tool,hydra-defect-prediction}, 
pattern mining \cite{pattern-infer-tosem-23,anti-pattern,mine-fix-patterns,fixminer},
automated program repair for patch generation \cite{LeGoues2012GenProg,ifixr, CURE-icse21, capgen-ICSE18,tbar,RewardRepair-icse22}, and patch assessment \cite{tina-tosem-patchassess,ODS,TianASE20,tian2021failtest}. 
Current bug datasets can be categorized into two main types. The first type involves human curation, resulting in carefully processed but small-scale datasets, such as ManyBugs and IntroClass \cite{LeGoues15tse}, Defects4J \cite{defects4j} and QuixBugs \cite{quixbugs-jss}. 
All of these datasets contain fewer than \numprint{1000} bugs each.
On the other hand, the second type comprises large-scale datasets mined from code repositories, such as ManySStubBs4J \cite{manystupidbugs}, which processes commit messages or issues with text processing. However, this approach tends to be imprecise in identifying bugs or classifying their types, relying on simple keyword comparison. 
Despite this loss of precision, crawling historical bug-fix commits from open-source projects remains a widely used approach to collect training bug-fix data in both academic \cite{CoCoNuT, SEQUENCER} and industry \cite{getafix-2019-oopsla, ml-repair2021-maps}, due to the ease with which it can construct large-scale datasets. 

There are several key challenges to collecting a truly useful, precise bug dataset: 

\begin{itemize}
\item \textbf{Problem 1 - Data quality and scalability:} Manual collection ensures the quality of the dataset but does not scale.  Automatic collection with keyword matching on commits is scalable, but imprecise~\cite{BugOrEnhancement-2008}.
\item \textbf{Problem 2 - Language diversity:} Most datasets consist only of a single language, limiting their usability and posing a threat to the external validity of software tasks.
\item \textbf{Problem 3 - Lack of metadata:} The most common metadata provided by the dataset per bug is a commit ID or commit message. Important information such as the type of bug, the severity of the bug, and the date of discovery is missing.
\item \textbf{Problem 4 - Lack of tests:} Only a handful of bug datasets provide test cases for the exposed bug. This challenges both bug reproduction, and validating bug fixes beyond the one provided by the developer.
\end{itemize}

In this paper, we introduce the \approach, a novel approach to tackle the imprecision problem in bug dataset creation. 
Our approach offers a curated collection of software defects, providing not only accurate bug type classification but also meta information to describe  the bug, including original buggy code, fix code, precise location, error type, and available executable/reproducible test cases. We denote this information as $<bug, fix, loc, type, (test)>$. \footnote{Not all the bug-fix data contains executable test cases.}

The \approach  is based on two components: \textit{bug tracker} for general bug collection, and \textit{bug injection} for project-specific bug generation. 
The bug tracker focuses on extracting bug-fix commits, leveraging available code repositories (e.g., GitHub, Bitbucket, Gitlab,  SVN) and external bug repositories (e.g., OSS-Fuzz, Jira, Apache Bug Report). 
The code repositories provide the exact source code changes made to address bugs, while the bug repositories offer detailed metadata about each bug. However, simply considering each repository in isolation is incomplete. 
Code repositories contain source code changes but lack clear bug metadata; bug repositories  do not always include information about the exact source code changes.
Therefore, we devise a method to merge these two sources, combining the advantages of both. 

Bug injection automatically generates artificial unseen bugs, each of which is specified by an existing test suite with at least one failing test that exposes the bug. 
To initiate the bug generation process, we begin with code extracted from various projects, ensuring that all existing tests pass successfully. The code noising tool is then employed to deliberately introduce changes into this code, simulating bug injection. It is essential to acknowledge that not all the injected code is genuinely buggy as some changes might be benign; we use the original passing tests to identify meaningful injected faults. The test failure diagnosis obtained via this validation process precisely identifies the type of bug introduced.

Considering the challenge objective, we consider bugs from both the bug tracker and bug injection components to be beneficial for industrial settings. The bug tracker ensures that we have access to a diverse range of bug fixes from real-world projects, making the dataset more representative. Bug injection allows us to have project-specific bugs that are difficult to learn from a general dataset, and tailored for industrial settings.

The two bug tracking sources that we use for the bug tracker are: the National Vulnerability Dataset (NVD) \footnote{https://nvd.nist.gov/vuln}  and OSS-Fuzz \footnote{https://google.github.io/oss-fuzz}.
NVD is a repository of vulnerability management data represented using the Security Content Automation Protocol (SCAP); it includes databases of security checklist references, software flaws, misconfigurations, product names, and impact metrics.
OSS-Fuzz, provided by Google, is an open source continuous fuzzing service. It uses random inputs (fuzzed data) to discover potential vulnerabilities, bugs, or crashes. OSS-Fuzz is specifically designed to identify security vulnerabilities and defects in open-source projects.
We implement the bug injector with 16 single-statement injection rules from previous work~\cite{selfAPR2022}. 

In the end, to date, the \approach collected a total of \numprint{1057818} bugs from \numprint{2968} open-source projects. We name this dataset 
\texttt{PreciseBugs}.
Out of these, \numprint{12602} bugs are contributed by the bug tracker (NVD and OSS-Fuzz), while   \numprint{1045216} bugs are generated by the bug injector. To our knowledge, this is the largest executable bug dataset with precise bug information compared with related work. 

To sum up, we make the  following contributions:
\begin{itemize}
\item  We introduce \approach for collecting a precise bug dataset to collect general-wise real-world bug-fix data and project-specific bug-fix data.
\item We present a comprehensive dataset named 
\texttt{PreciseBugs}, which includes \numprint{1057818} bugs based on \numprint{2968} open-source projects across more than six programming languages.
\item We make our dataset readily accessible and openly available to the community through the link: \color{blue}{https://github.com/SophieHYe/PreciseBugs}.

\end{itemize}


\section{Background}
\label{sec-backgound}

In this section, we give background on bug-fix datasets in the program repair literature.
\autoref{background-dataset} presents a summary of 20 extensively utilized bug-fix datasets. These bug-fix dataset collection approaches (found in the seventh column of \autoref{background-dataset}) consists primarily of four commonly used methods. We explain them according to the collection approaches in the following.

\underline{Commits+Text Processing}: Most datasets use commit mining and filtering against specific keywords (e.g., BugsJS \cite{bugsjs-icst19}) to approximate bug-fix commits. Commonly used keywords include ``fix'' and ``bug".

\underline{Commits+Continuous Integration (CI)}: Another approach uses continuous integration (CI) infrastructure for bug collection, as observed in projects such as Bears \cite{Bears} and Bugswarm~\cite{bugswarm-icse19}. This method revolves around checking the CI status of two consecutive commits. When the first commit fails but the second one passes, it is considered a bug-fix pair. One of the advantages of this approach is gaining additional insights into the failing tests that triggered the CI build failure, and thus the bug. 

\underline{Programming Contests}:
This category of research involves collecting bug-fix commits from contests or programming competitions, as demonstrated by projects such as CodeFlaw~\cite{Codeflaws} and QuixBugs~\cite{quixbugs}. In programming contests, developers are presented with a problem description and test case specifications, which facilitates bug-fix collection encompassing both bug types and corresponding failing tests. The bug-fix commits are obtained by analyzing users' submission histories and identifying two submissions where the first submission fails and the second one is accepted. However, this approach is limited by the scarcity of available contests and the relatively low participation of developers.

\underline{Student Assignment Submissions}:
This group of approaches identifies and collects bug-fix commits from student submissions in introductory programming courses, exemplified by projects like IntroClass~\cite{LeGoues15tse} and Refactor~\cite{refactory-ase19}. The process of collecting bug-fix commits in this context is similar to that for programming contests, analyzing student submission history. However, it is important to note that this approach is limited to collecting bug-fix commits from relatively small programs, typical of introductory programming assignments.

By considering different data collection approaches, we make the following implications.

\textbf{Problem 1 - Data quality and scalability:} It becomes evident that manual validation for bug-fix collection is not scalable. While manual validation ensures the quality of the collected bugs by including precise bug types and regression test cases, it requires extensive and laborious efforts in committing searches, data cleaning, and execution. As a result, manual bug collection is impractical for collecting large-scale datasets. For instance, even widely-used datasets like Defects4J \cite{defects4j} and ManyBugs \cite{LeGoues15tse} are limited to containing fewer than a thousand bugs each. On the other hand, most of the datasets in \autoref{background-dataset} use commits+text processing to crawl bugs. Although this approach is scalable and can collect millions of examples. \citeauthor{BugOrEnhancement-2008} have shown that text classification is not enough to classify the intent of a commit \cite{BugOrEnhancement-2008}.

\textbf{Problem 2 - Language diversity:} There is a lack of language diversity in the bug dataset. Of the 20 widely used bug-fix datasets, the majority (15 out of 20) are focused solely on a single programming language, leaving only 5 datasets that cater to multiple languages. This language bias in bug datasets can limit the generalizability of research findings and may not adequately represent the diverse landscape of software development. Different programming languages have unique syntax, semantics, and coding practices, leading to varying types of bugs and bug-fix patterns. Therefore, incorporating multiple programming languages in bug-fix datasets is crucial to enable a more comprehensive understanding of bugs and their repairs across different language ecosystems.

\textbf{Problem 3 - Lack of metadata:} A significant issue with many bug datasets is the lack of sufficient metadata on the bugs. Crucial information, such as the date of bug discovery, bug type, commit author, and bug severity, is often missing. This absence of metadata poses challenges in analyzing the bug dataset beyond its initial intended use case. Proper metadata is essential for conducting in-depth research and understanding the characteristics of bugs, their patterns, and the context in which they occur.

\textbf{Problem 4 - Lack of tests:} Only a few bug datasets come with accompanying tests. Test cases are an indispensable means of specifying program correctness and validating bug-fixes. Having test cases allows researchers to verify the correctness of bug-fixes independently from the developer bug-fix present in the dataset. Moreover, test cases enable various dynamic analyses on the source code, such as fault localization-based on test coverage. Including test cases in bug datasets enhances the overall usability and utility of the dataset, enabling researchers to conduct more extensive and accurate evaluations of program repair techniques.

In our PreciseBug dataset, we strive to address these four challenges, with the ultimate goal of creating a comprehensive, precise, and large-scale bug dataset. The methodology used to construct this dataset is detailed in \autoref{sec-pricesebug}.

\begin{figure}[t!]
\includegraphics[width=.5\textwidth]{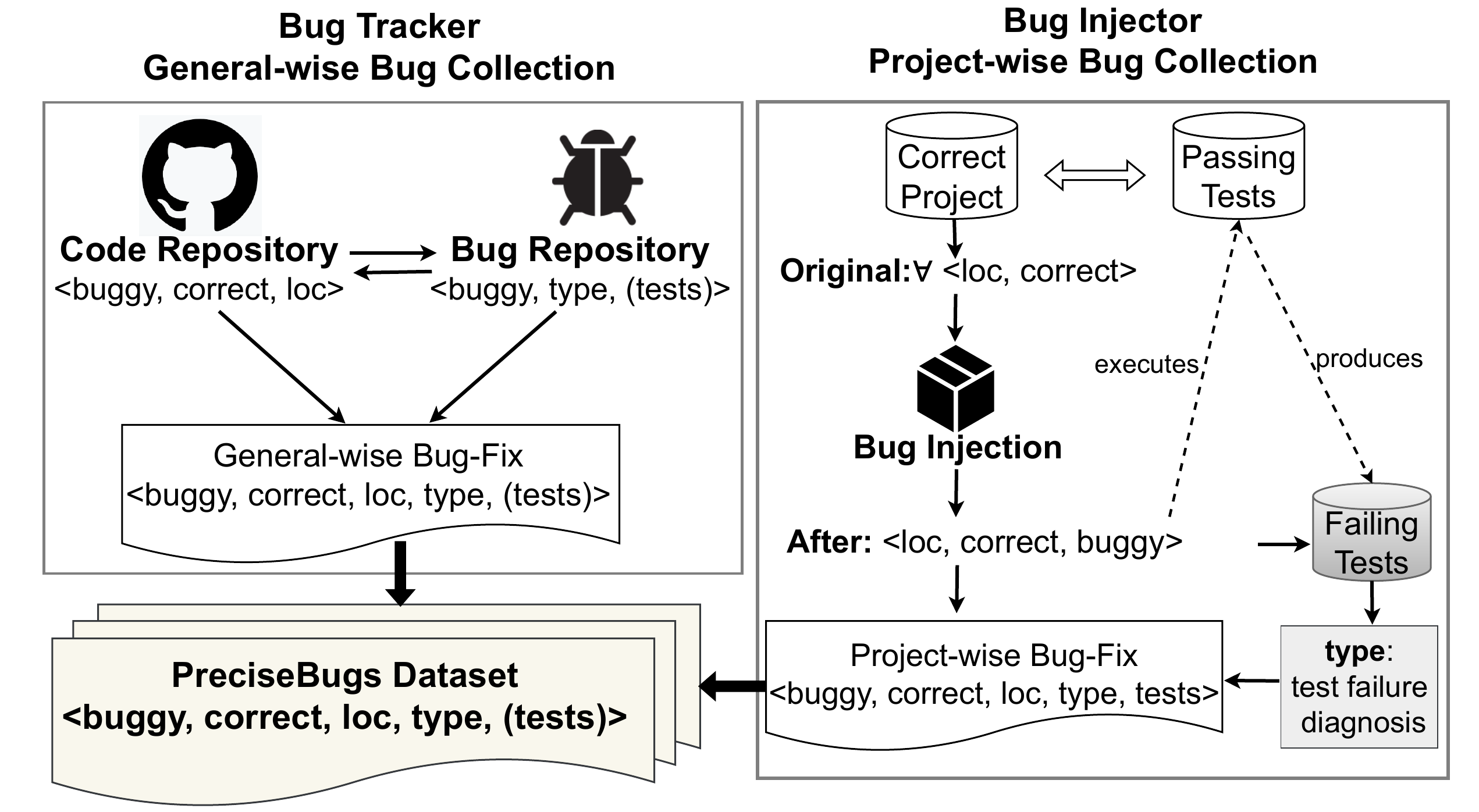} 

 \caption{An Overview of PreciseBugCollector.}
\label{fig:PreciseBugCollector-overview}
\end{figure}

\begin{figure*}[th]
\includegraphics[width=0.98\textwidth]{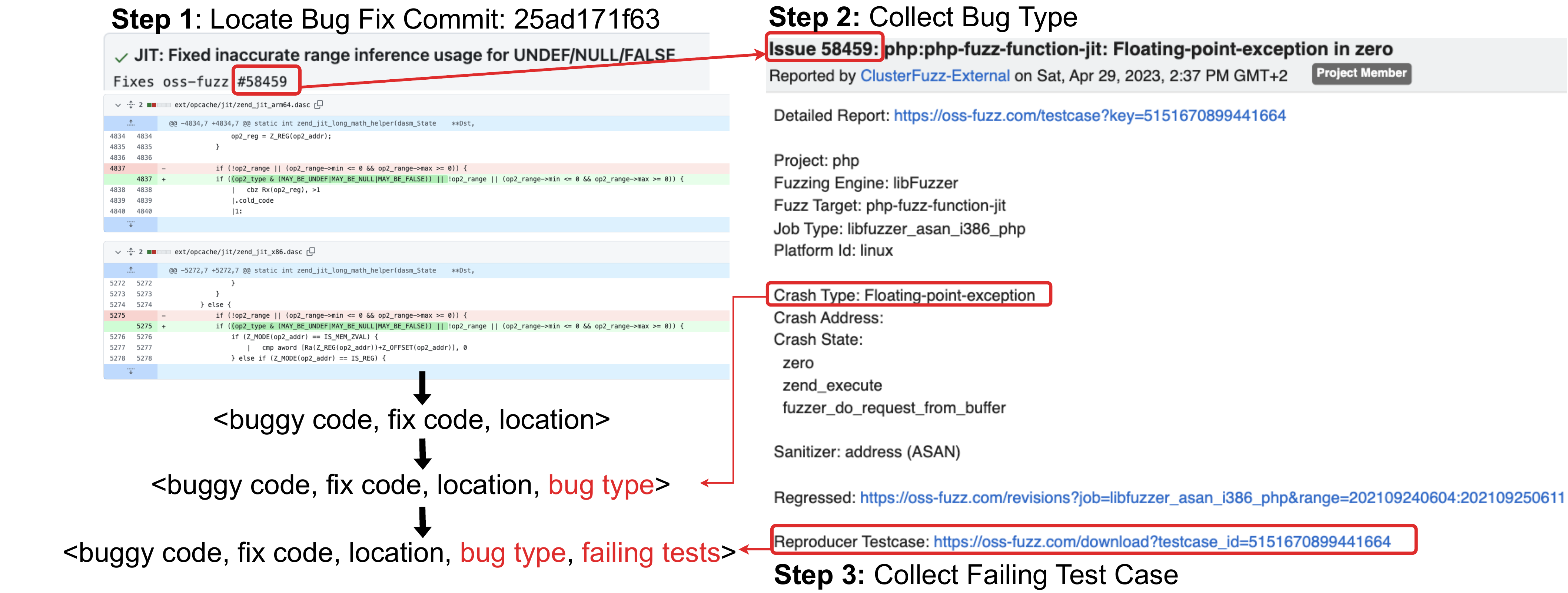} 
\centering
 \caption{Running example of the bug tracker component with the external bug repository OSS-Fuzz. }
\label{fig-runningexample}
\end{figure*}

\section{PreciseBugCollector}
\label{sec-pricesebug}

\autoref{fig:PreciseBugCollector-overview} gives an overview of the \approach collection architecture. \approach consists of two novel components: a bug tracker for general bug collection and bug injection for project-specific bug collection.

The bug tracker creates real-world bug datasets by establishing connections between code repositories and external bug repositories. We use GitHub as the code repository instance, given its vast size, hosting over 28 million public repositories~\cite{github-code-repository-msr}.
For bug repositories, many external bug repositories can be used, so long as they provide precise bug types and providing unique bug identifications that can be linked to code repositories (like an issue ID).  Widely-used repositories that meet these criteria include the Common Vulnerabilities and Exposures (CVE) repository by the NVD, OSS-Fuzz, and Apache bug repositories.

The bug injection component relies on bug injection tools to generate a project specific bug-fix dataset. Bug injection generates new bugs by introducing noise into   code that passes all tests in its accompanying test suite, and using those tests to evaluate the bugginess of the modified noisy code. The test suite not only plays a role in determining the bugginess of the noised code but also provides essential test failure diagnosis information, describing the bug.

\subsection{Bug tracker: mapping code and bug repositories}

\approach leverages external bug repositories to acquire accurate bug type information. 
Given a commit from a code repository, one can easily obtain the buggy code and fix code and their diffs to extract the code change locations, denoted as $<buggy, fix, loc> $. However, as discussed in \autoref{sec-backgound},  it is challenging to summarize bug types from fix commit messages. 
Our work introduces a novel approach to establishing mappings between code and bug repositories, revealing the actual bug type, and producing $<buggy, fix, loc, type>$ information.
Moreover, there exist bug repositories that also contain failing tests that expose the bugs; this is crucial  for buggy code reproduction and execution~\cite{importance-reproduce-bugs}.
Consequently, the bug tracker may obtain the bug-fix dataset in the format of $<buggy, fix, loc, type, (tests)>$, where the test information is optional depending on the external bug repositories. 

We use two external bug repositories: NVD and OSS-Fuzz, both of which assign unique identifiers to vulnerabilities and bugs (a CVE in the NVD, and unique identifiers in OSS-Fuzz). Both NVD and OSS-Fuzz are large, with 220,748 CVE records in the NVD, and 28,000 bugs in OSS-Fuzz. 
Notably, both repositories support different programming languages, including C/C++, Rust, Go, Python, and Java/JVM. 

\textbf{CVE collection-based on NVD.}
NVD serves as an invaluable resource for mining verified vulnerabilities reported by humans. Many important and well-known vulnerabilities are reported there, including HeartBleed (CVE-2014-0160\footnote{\url{https://nvd.nist.gov/vuln/detail/cve-2014-0160}}), Meltdown, (CVE-2017-5754\footnote{\url{https://nvd.nist.gov/vuln/detail/CVE-2017-5754}}), Spectr (CVE-2017-5753 and CVE-2017-5715 \footnote{\url{https://nvd.nist.gov/vuln/detail/cve-2017-5753} and \url{https://nvd.nist.gov/vuln/detail/cve-2017-5715}}), and Log4Shell (CVE-2021-44228\footnote{\url{https://nvd.nist.gov/vuln/detail/cve-2021-44228}}).
Each vulnerability reported on NVD includes essential the vulnerability description, its type identified by the CWE (Common Weakness Enumeration) ID, a severity level, and references. References often contain patches to the vulnerabilities, providing source code before and after the vulnerability fix. Our systematic procedure for collecting vulnerabilities from NVD is as follows:

\begin{enumerate}
\item We use the NVD API~\footnote{https://nvd.nist.gov/developers/vulnerabilities} to download the complete vulnerability metadata, a total of \numprint{217403} vulnerabilities.
\item We filter vulnerabilities by identifying external links that lead to GitHub commits and with the \textit{Patch} tag. This narrows the vulnerabilities to 9,759.
\item We extract the fixed source code from the corresponding GitHub commits. Extraction may fail due to name changes or repositories having been removed.
\end{enumerate}

After the last step, we are left with a dataset of 8487 vulnerabilities, each accompanied by its metadata and the corresponding vulnerability fix.
The decision to extract vulnerability fixes exclusively from GitHub commits is based on the fact that patches reported to NVD come in various formats, making them challenging to parse. These formats may include links to patches on patched product download sites, blog websites, and bug discussions on various platforms, among others. By focusing on GitHub commits specifically tagged as \textit{Patch}, we can mitigate noise and ensure a more consistent and reliable dataset. Additionally, using GitHub allows us to easily retrieve the project name from the repository name, further enhancing accuracy.

\textbf{Bugs collection-based on OSS-Fuzz.}
Fuzz testing \cite{fuzzing-2018, fuzz-19, fuzzing-20} is a widely acknowledged technique for detecting programming errors, especially critical issues like buffer overflows.
In contrast to the vulnerabilities collected in NVD, each bug found by OSS-Fuzz provides explicit failing tests that expose the bugs.
Our systematic procedure of collecting bugs from OSS-Fuzz is as follows:
\begin{enumerate}
\item We conduct an exploration of the OSS-Fuzz repository, and retrieve open-source projects that have registered to use the OSS-Fuzz infrastructure, resulting  in a total of 28348 bugs from 557 open source projects.
\item To obtain the bug-fixes for each project, we first filter GitHub commit messages and search for OSS-Fuzz identifiers. Then, we use these identifiers to query the OSS-Fuzz API, which retrieves crash types per bug. After this step, we are left with  a dataset of 4025 bugs collected from six programming languages.
\item (optional) We gather the failing tests that expose the bugs from OSS-Fuzz. However, not all projects support the download of failing test cases due to authorization issues.
\item (optional) We create reproducible bugs by executing collected failing tests against the fixed code (i.e., the patched program) to check if they pass. We execute the same tests against the buggy program to verify that it indeed results in test failures.

\end{enumerate}

\begin{table}[t!]
\renewcommand{\arraystretch}{1.0}
\footnotesize
\centering
\caption{Injection rules for bug creation.}

\begin{tabular}
{llccccccccccc}
\hline
Bug Injection Rules & Description \\
\hline
Rule-1 & modify declaring type \\

Rule-2 & modify operator & \\

Rule-3 & modify literal &    \\
Rule-4 & modify constructor &    \\

Rule-5 & swap argument &    \\

Rule-6 & modify boolean expression &    \\
Rule-7 & modify invocation &    \\
Rule-8 & compound modification &    \\
Rule-9 &  replace similar  statement \\
Rule-10 &  move  statement \\
Rule-11 &  insert  statement \\
Rule-12 &  wrap  statement \\
Rule-13 &  insert  block \\
Rule-14 &  delete  block \\
Rule-15 &  unwrap  block \\
Rule-16 &  remove  block \\
\hline

\end{tabular}

\label{injection-rules}
\end{table}

\textbf{Running Example of bug tracker with OSS-Fuzz}
Figure \ref{fig-runningexample}  illustrates the process of collecting bugs from the PHP project using the external bug repository OSS-Fuzz. 
In the initial step, we browse the PHP GitHub commits to collect the OSS-Fuzz issue identifier. Next, we establish a link between the GitHub code repository and the OSS-Fuzz bug repository to identify the crash type (e.g., floating point exception) and obtain the corresponding failing test cases. Finally, we assess the executability and reproducibility of the collected bugs by separately executing the fix code and buggy code against the test cases.

\subsection{Bug Injection: Creating Project-specific Bugs}
Bug injection involves deliberately introducing faults into a stable codebase and is widely used to test software reliability and dependability \cite{fault-injection-1997,fault-injection-2022}. Bug injection is heavily used in mutation testing to test the strength of existing test suites~\cite{jia2010analysis}, as well as to test bug finding tools~\cite{dolan2016lava,pewny2016evilcoder,roy2018bug}. Recent work in automatic program repair with machine learning also uses bug injection to create training data to train deep learning program repair models~\cite{allamanis2021self-buglab, Yasunaga20DrRepair}.

\looseness-1
We adopt this approach to generate project-specific bugs by intentionally corrupting the correct codebase and execute injected bugs against existing  test suites to collect precise bug type from compiler and test failure diagnosis.
Project-specific bugs hold significant value in the industrial context, particularly for companies working on projects that require domain-specific knowledge and adhere to unique coding styles. These specialized bugs are challenging to learn from general-purpose datasets.

\begin{table*}
\renewcommand{\arraystretch}{1.5}
\small
\caption{An overview of bug-fix data collected by \approach.}

\resizebox{\textwidth}{!}{\begin{tabular}
{l|rr|rr|rrr|rr}
\hline
 Languages & \multicolumn{2}{c}{    NVD-based Vulnerabilities} &  \multicolumn{2}{c}{ OSS-Fuzz-based Bugs  }&  \multicolumn{3}{c}{Injection-based Bugs } &  \multicolumn{2}{c}{Summary} \\
& \#projects & \#behavior bugs & \#projects & \#behavior  bugs & \#projects & \#compilation bugs & \#behavior bugs & \#projects &  \#bugs \\
\hline

C  & 419 & 2559 &  12 & 1604 &  1 & 1735 & 637 & 424 & \numprint{6535} \\
C++ & 117 & 617 & 53 & 2439  & - & -  & - & 151 & 3056 \\
Python & 291 & 468 & 3 & 3 & 3 & 4372 & 753 & 292 & 1224  \\
Java &199 & 309 &10 & 48  &  17 & \numprint{631015} & \numprint{409230} & 213  & \numprint{1040602} \\
Go & 189 & 323& 9 & 15 & -& -& -  & 195  & 338 \\
Rust &28 & 38&5 & 6 & - & -& - & 30 & 44 \\
PHP & 597 & 1846 &- & -  & -  & -  & - & 597 & 1846   \\

Others & 1065 & 4173   & - & - & -  & -  & - & 1065 & 4173  \\
\hline
All &2905 & 8487    &  79 & 4115 & 21 & \numprint{637122} &  \numprint{410620} & \colorbox{gray!50}{\numprint{2968}}&   \colorbox{gray!50}{\numprint{1057818}}\\


\hline

\hline

\end{tabular}}

\label{table-precise-bugs}
\end{table*}

\begin{algorithm}[t!]
\footnotesize
  \caption{Injection-based Bug Creation}
  \begin{algorithmic}[1]
 \STATE \textbf{Input:} a correct program $CorrectCodebase$, Injection rules $rules$, test suite $tests$
 \STATE PreciseBugs $\gets \emptyset$  
 \FOR{loc, fix in $CorrectCodebase$} 
    \STATE  $ type$ $\gets$ apply(rules, loc, fix)
    \STATE $error, type$  $\gets$ compile($buggy$)
    \IF{ !$error$}
    \STATE $type$  $\gets$ execute($buggy$, $tests$)
    \ENDIF
    \IF{$type$}
         \STATE PreciseBugs $\gets$ $<$  $buggy$, $fix$, $loc$, $type$, $tests$ $>$
    \ENDIF
 \ENDFOR
  \RETURN PreciseBugs
  \end{algorithmic}
  \label{alg:code-injection}
\end{algorithm}

\emph{Input and Output.}
Algorithm \ref{alg:code-injection} illustrates the bug injection  process.
Bug injection takes a "correct" project codebase and its corresponding tests as input, where "correct" means that the considered project passes all test cases. 
For every statement present in the project codebase, bug injection introduces noise to that statement with the aim of changing the program's execution behavior (see line 3  in Algorithm \ref{alg:code-injection}). The original test suite is then utilized to assess whether the injected code's behavior has changed by yielding at least one failing test.

We first attempt to compile noised code. 
If any compilation error message is produced,  this injected noisy code is deemed to cause a compilation error (line 5). Otherwise, the noisy code is further executed against the test suite to determine whether it causes a behavioral bug  (line 6 and line 7).
If any tests fail, a test failure diagnosis is performed (line 9 and line 10).  Otherwise, it is discarded.

\emph{Injection Rules.}
We employ abstract syntax tree  (AST)-based 16 bug injection rules from prior work~ \cite{selfAPR2022}. 
\autoref{injection-rules} details the rules, which cover different granularities of code transformation, such as type, operation, literal, valuables, expression, statement, and block. Note that these injection rules provide more diverse code transformation rules than existing works that  mostly focus on operators and variables \cite{mutationtesting,major,patra2021semantic,allamanis2021self-buglab}.

\subsection{Comparison of Collected Bugs}

\begin{table}[h!]
\renewcommand{\arraystretch}{0.5}
\centering
\footnotesize
\caption{Comparisons of different sources considered by  \approach.}

\resizebox{0.5\textwidth}{!}{\begin{tabular}
{lcccccccccccc}

\hline
\multirow{2}{*}{\thead{Sources}}  & \multirow{2}{*}{  \thead{Compilation \\ Bug}} & \multirow{2}{*}{\thead{Behavior\\ Bug}} & \multirow{2}{*}{ \thead{Fix}}  & \multirow{2}{*}{ \thead{Location}}  & \multirow{2}{*}{ \thead{Type}}   & \multicolumn{2}{c}{ \thead{Failing Tests}} \\

&&&&&&\thead{New}&\thead{Existing}\\

\hline

 \thead{CVE collection-based \\on NVD} & 
  \colorbox{red}{\xmark}&
   \colorbox{green}{\checkmark} & 
  \colorbox{green}{\checkmark} & \colorbox{green}{\checkmark}  & \colorbox{green}{\checkmark} & \colorbox{red}{\xmark} & \colorbox{red}{\xmark} \\

\hline

 \thead{Bugs collection-based \\on OSS-Fuzz} & 
 \colorbox{red}{\xmark}&
  \colorbox{green}{\checkmark} & 
   \colorbox{green}{\checkmark} & \colorbox{green}{\checkmark}  & \colorbox{green}{\checkmark} & \colorbox{green}{\checkmark}  & \colorbox{red}{\xmark} \\

\hline

 \thead{Bugs based \\on Injection} & 
\colorbox{green}{\checkmark} & 
\colorbox{green}{\checkmark} & 
 \colorbox{green}{\checkmark} & 
 \colorbox{green}{\checkmark} & \colorbox{green}{\checkmark}    & \colorbox{red}{\xmark} & \colorbox{green}{\checkmark}\\
\hline

\end{tabular}}

\label{approach-comparison}
\end{table}

\autoref{approach-comparison} provides a comparison of the three considered bug collection approaches. All three approaches gather bug information in the format of $<bug, fix, location, type>$. Both bug tracker with OSS-Fuzz and bug injection approaches can collect reproducible bugs along with their corresponding failing tests.
However, the failing tests from OSS-Fuzz include additional new test cases that go beyond the existing test suite. Failing tests from the bug injection are the existing test cases in the test suite that were originally passed before the bug injection took place.

Notably, Bug injection is the only one that produces compilation bugs, where the error type and error message directly come from the compilers. Therefore, bug injection is a powerful approach to constructing both compilation and behavior bugs.

\section{Evaluation}

To evaluate the \approach, we propose the three following research questions:

\begin{itemize}
\item  \textbf{RQ1:}  What is the effectiveness of the \approach to construct bug-fix commits?
\item \textbf{RQ2:} What is the distribution of bug types collected by \approach?
\item \textbf{RQ3:}  What is the distribution of the time period covered by bugs collected by \approach?

\end{itemize}

\subsection{ RQ1: Number of Bugs}

\emph{Methodology for RQ1.}
We summarize the number of bugs that \approach has collected. Specifically, we analyze 1) the number of major programming languages, 2) the number of projects, and 3)  the number of bugs  that \approach is able to cover.

\emph{Result for RQ1.}
\autoref{table-precise-bugs} gives an overview of the bug-fix data obtained by \approach.
\approach collected 8487 CVEs from 2905 projects-based on NVD dataset, 4115 bugs from 79 projects-based on OSS-Fuzz bug repository, and respectively 637122 compilation bugs and 410620 behavior bugs from  21 open source projects.

In total, \approach obtained \numprint{1057818} bugs from  \numprint{2968} projects and more than six programming languages. To our knowledge, this is the largest  bug-fix collection with precise bug type and execution information to date. All three are able to collect  thousands of bugs, and each of them individually collects more than the largest prior dataset shown in \autoref{background-dataset}.
\autoref{lst:three-examples} gives three examples of collected bug-fix data from each approach.  
Now, we discuss the implications of these results.

\emph{CVEs from NVD cover many projects.}
We  have collected CVEs from 2905 open source projects by tracking to NVD, many more than the 79 and 21 projects covered by the other two approaches. This discrepancy is primarily due to the NVD's establishment as one of the earliest and widely utilized bug repositories, dating back to 2004 (whereas OSS-Fuzz emerged in 2016).

\emph{Bug injection generates the most bugs.}
The bug injection approach, while covering the fewest  projects, proves to be highly effective in generating a large number of bugs. This is primarily attributed to the project-specific nature of the bug injection technique, which aims to traverse every statement in the program. However, it is worth noting that the execution cost for the bug injection is relatively high, which is why we restricted our experimentation to only 21 projects from three languages.

Bug injection generates various numbers of bugs for different projects despite the same code corruption rules. 
This is because the number of bugs generated relies on the number of lines of code (LOC) and test suite size and strength. 
In our experiment, the considered 17 Java projects (e.g., Closure,  JacksonDatabind, etc.) are comprise 25,000 LOC and more than 2000  test cases, meaning that more bugs are generated compared with projects in C and Python.

\begin{figure*}[t!]
     \centering
     \begin{subfigure}[b]{0.32\textwidth}
         \centering
         \includegraphics[width=\textwidth, height=4.7cm]{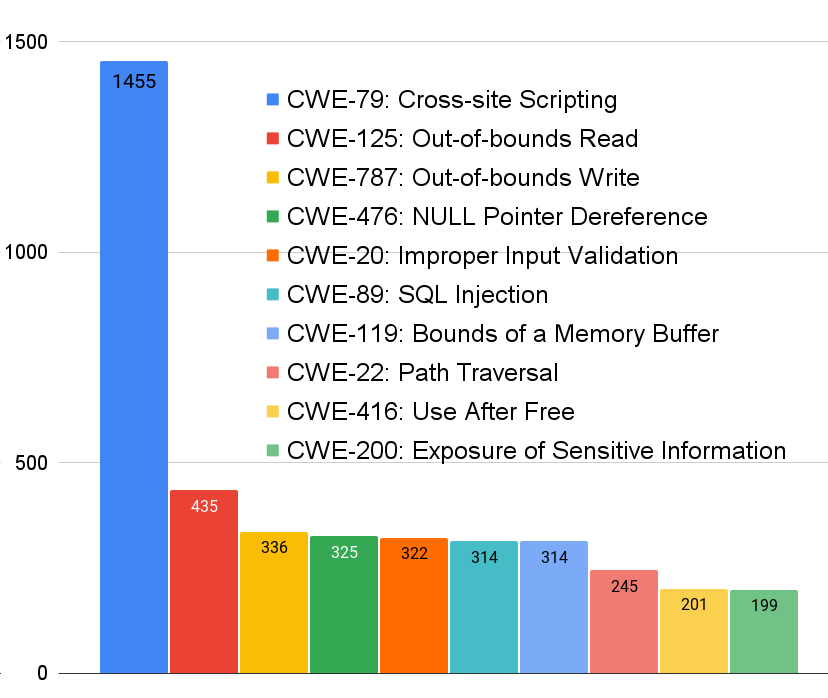}
         \caption{NVD-based CVE Type}
         \label{fig:y equals x}
     \end{subfigure}
     \hfill
     \begin{subfigure}[b]{0.32\textwidth}
         \centering
         \includegraphics[width=\textwidth, height=4.7cm]{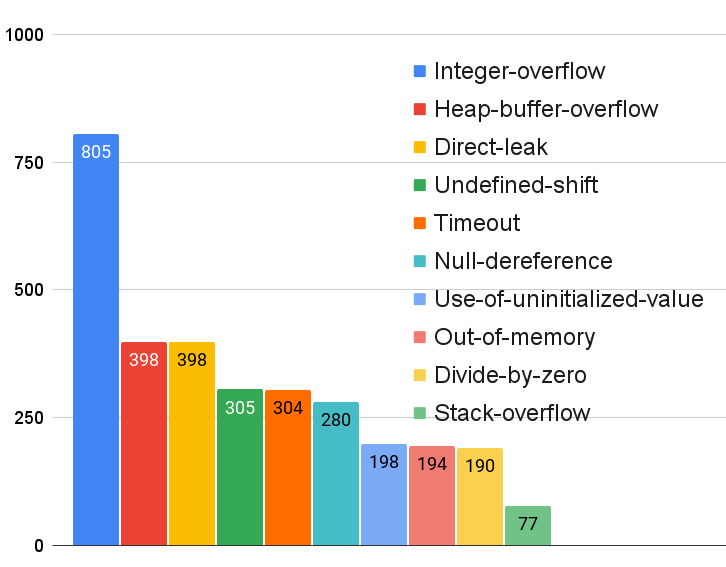}
         \caption{OSS-Fuzz-based Bug Type}
         \label{fig:three sin x}
     \end{subfigure}
     \hfill
     \begin{subfigure}[b]{0.33\textwidth}
         \centering
         \includegraphics[width=\textwidth,height=4.7cm]{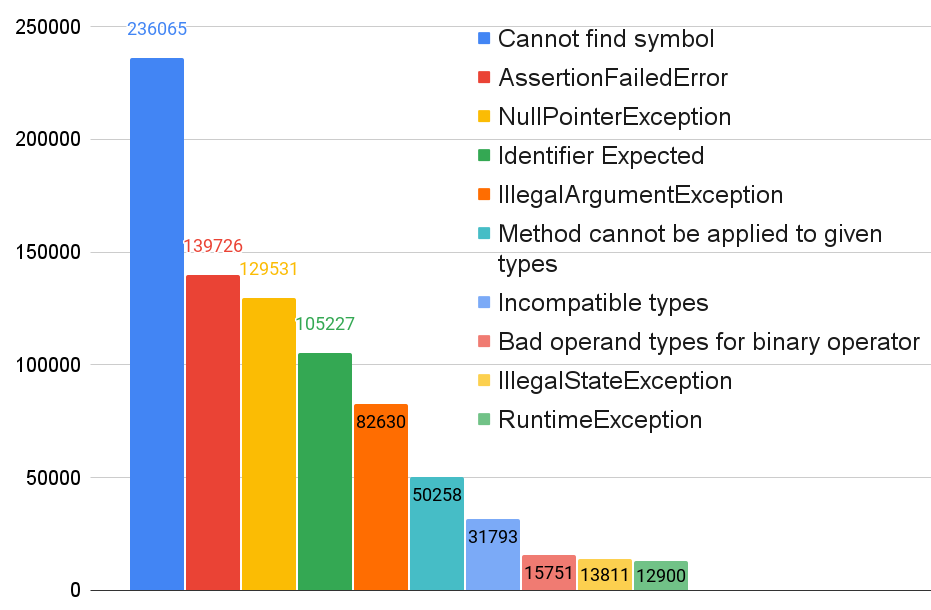}
         \caption{Injection-based Bug Type}
         \label{fig:five over x}
     \end{subfigure}
        \caption{Top-10 bug types in three considered sources by \approach.}
        \label{fig:three-types}
\end{figure*}

\begin{mdframed}[nobreak=true]
Answer  to  RQ1: 
\approach  gathered a  total of 1,057,818 precise bugs from 2,968 open source projects.
\end{mdframed}

\begin{listing}[t!]
\noindent\begin{minipage}[b]{.9\columnwidth}
    \begin{lstlisting} [firstnumber=366] 

<@\colorbox{red!60}{366 - if(length<0xffffffff \&\& length+index < size())\hspace{.5mm}}@> 
<@\colorbox{green!60!}{366 + if(length < size() - index) \quad\quad \quad\quad\quad\quad \quad\quad \quad\quad \quad    }@>
<@\textbf{CVE ID: CVE-2012-1584}@> 
<@\textbf{Type: CWE-189 Numeric Errors}@>
<@\textbf{Loc: Before:  366, After:  366}@>



    \end{lstlisting}
 
     \subcaption{An  NVD-based vulnerability collected  from project TagLib.}
   \label{motivate-human-patch}  
    \end{minipage}%
    \hfill
    \begin{minipage}[b]{.9\columnwidth}
    \begin{lstlisting}[firstnumber=419] 
<@\colorbox{red!60}{2859 - buf[i + 0] = - (USE\_FIXED + 1)*buf[i + 0];\hspace{6.3mm}}@>  
<@\colorbox{red!60}{2859 - buf[i + 1] =  (USE\_FIXED + 1)*buf[i + 1];\hspace{9.2mm}}@>
<@\colorbox{green!60!}{2859  + buf[i + 0] = -(int)(USE\_FIXED+1U)*buf[i + 0];\hspace{1.9mm}}@>  
<@\colorbox{green!60!}{2860  + buf[i + 1] =    (int)(USE\_FIXED + 1U)*buf[i + 1];\hspace{0.4mm}}@>  

<@\textbf{OSS-Fuzz Issue ID: 57986}@> 
<@\textbf{Type: Integer-overflow}@>
<@\textbf{Loc: Before:  (2859,2860), After:  (2859,2860)}@>


    \end{lstlisting}
 
    \subcaption{An OSS-Fuzz-based bug collected  from project FFmpeg. }
 \label{motivate-noncompile-patch}   
\end{minipage}%
\hfill
    \begin{minipage}[b]{.9\columnwidth}
    \begin{lstlisting}[firstnumber=419] 
<@\colorbox{red!60!}{4493 -  Collection c = r.getAnnotations();\hspace{18.1mm}}@> 
<@\colorbox{green!60!}{4493 +  if (r != null) \{\hspace{44.9mm}}@>   
<@\colorbox{green!60!}{4494 +   Collection c = r.getAnnotations();\hspace{18.2mm}}@> 
<@\colorbox{green!60!}{4495 +   \}\hspace{67.mm}}@> 
<@\textbf{Failing Test: LogAxisTests:testXYAutoRange1}@> 
<@\textbf{Type: NullPointerException}@>
<@\textbf{Loc: Before:  4493, After:  (4493,4495)}@>



    \end{lstlisting} 
    \subcaption{An  injection-based bug collected from project JFreeChart.}
    \label{motivate-correct-patch} 
\end{minipage}%
\caption{Examples of collected bugs from three sources.}
\label{lst:three-examples}
\end{listing}

\subsection{RQ2: Types of Bugs}

\emph{Methodology for RQ2.}
In this RQ, we investigate  the type of collected bugs. Specifically, we look at unique bug types that each component brings to \approach.

\emph{Result for RQ2.}
\autoref{fig:three-types} gives the top-10 bug types from each component of \approach. The distribution of bug types varies by component.  This is due to the differences among three sources: CVE types are labeled by humans, OSS-Fuzz types are labeled by fuzzing tests, and injection-based bug types are extracted by test failure diagnosis.

\emph{CVE-based Types.}
Collected CVE bug-fix commits contain CWE IDs, a community-developed list of software and hardware weakness types.
The most common CVE type in \approach is CWE-79\footnote{https://cwe.mitre.org/data/definitions/79.html}:  \texttt{Improper Neutralization of Input During Web Page Generation ('Cross-site Scripting')}, which accounts for 17.1\% (1455/8487) of the total CVE types. The remaining top-2 bug types are out-of-the-bounds read and out-of-the-bounds write.

\emph{OSS-Fuzz-based Types.}
For each collected OSS-Fuzz-based bug-fix commit, a unique issue ID leads to a certain crash type produced by fuzzing tests.
The dominant bug type from OSS-Fuzz is  integer-overflow, which  accounts for 19.6 (805/4115) of the total OSS-Fuzz-based collected bugs.
The remaining top-2 bug types are heap-buffer-overflow  and direct-leak, which are not included in the top 10 of  another two sources.

\emph{Injection-based Types.}
Each collected injection-based bug-fix commit is accompanied by a precise error message. Either this message comes from a compiler or a test suite execution result. The most frequent compilation error type from injection-based bugs is ``cannot find symbols'', while the top-2   behavior bug types are assertion failures with concrete error messages and null pointer exceptions.

\begin{mdframed}[nobreak=true]
Answer to RQ2: 
\approach contains a bug-fix dataset with  diverse and precise bug types, and each source contributes unique types of bugs. 

\end{mdframed}

\begin{table}[t!]
\renewcommand{\arraystretch}{1.28}
\footnotesize
\centering
\caption{Year distribution of collected bug-fix dataset. The \colorbox{gray!20}{rows} is valuable to avoid data leakage  evaluation for LLM models.}

\begin{tabular}
{lrrrrccccccc}
\hline
Year  &  NVD-based & OSS-Fuzz-based & Injected-based & Summary  \\
\hline
\rowcolor{gray!20}
2023&717&161& \numprint{1047724} & \numprint{1048620}    \\
\rowcolor{gray!20}
2022&1941&435&  -   & 2376  \\
2021&1399&521& - & 1920 \\
2020&865&413& - & 1278 \\
2019&662&495& - & 1157  \\
2018&644&1033& - &1677 \\
2017&723&1054&- & 1777\\
2016&520&3& -&  523  \\
$\leq$ 2015&1016&- & - & 1016 \\
\hline

\end{tabular}

\label{time-range}
\end{table}
\subsection{RQ3: Time Period and Data Leakage}

\emph{Methodology for RQ3.}
Large language models (LLMs) are evaluated on many existing bug datasets~\cite{alpha-repair,xia-llm-icse23}. Yet, these LLMs are also trained on data available prior to 2022 on the internet, which poses a threat of data leakage~\cite{tian2023chatgpt}. We analyze the \approach bug-fix dataset by year and particularly look at the bug-fixes available as of
2022 and 2023.

\emph{Result for RQ3.}
\autoref{time-range} summarizes the bug-fixes mined by \approach over timesince  2015. CVEs have a broader time distribution, as it is one of the  earliest and widely used vulnerability datasets. OSS-Fuzz began in 2016, and the majority of its bugs are repaired since 2017.
Injection-based bugs by their  nature are not available until the bug is injected, therefore, all the bugs are unseen by LLMs. 
Compared to related work, \approach provides an up-to-date bug-fix dataset.
The rows highlighted in \autoref{time-range} are useful to avoid data leakage in state-of-the-art software engineering maintenance tasks (bug detection, fault localization, program repair, etc).

\begin{mdframed}[nobreak=true]
Answer  to  RQ3: 
\approach contains valuable bug-fixes mined from 2022 and 2023, unseen by modern LLMs. These are valuable datasets for both training and testing to avoid data leakage.
\end{mdframed}

\section{Related Work}

\subsection{Bug-fix Dataset}

Now we present all the related work that collected a bug dataset.

Defects4J is a collection of 835 real-world Java programs with known software defects \cite{defects4j}. The bugs are curated from well known open source projects such as Apache Lang and Mockito.

ManyBugs and IntroClass datasets are both bug datasets focusing on C programs \cite{LeGoues15tse}. ManyBugs contains 185 defects collected from version control repositories of 9 projects. IntroClass consists of 998 defects from student written assignments for an introductory C programming course. All defects are reproducible with corresponding test cases.

Codeflaws is another bug dataset focusing on C programs \cite{Codeflaws}. It aims to fix the diversity and size problem of ManyBugs and IntroClass dataset. It crawls Codefroces for rejected submissions and finds another accepted submission by the same user for the same programming problem.

QuixBugs is a multi-lingual bug dataset, with 40 programs translated to Python and Java \cite{quixbugs}. All bugs are in one line of code, and each bug has passing and failing test cases.

CodRep is a machine learning on source code competition \cite{chen2018codrep}. The goal of this competition is to find the line number in a file to insert a given code line. All data are extracted from real-world one line commits.

\citeauthor{liu2018closer} collected 16450 bugs to do a systematic and fine-grained study to gain insights into tuning automatic program repair tools \cite{liu2018closer}. The bugs are collected from 6 Java projects using keyword matching and bug linking.

Bears is an extensible bug benchmark for program repair tools \cite{Bears}. It collects bugs in a unique way by looking at the commit building state from continuous integration to find potential bug-fix commits.

BugsJS is a JavaScript bug benchmark that can be used to facilitate the research of fault localization \cite{bugsjs-icst19}. All bugs are extracted from the issue tracking system of the selected popular repositories from GitHub.

BugSwarm, similar to Bears, is an extensible bug benchmark by mining bugs from continuous integration \cite{bugswarm-icse19}. The collected bugs are in two languages, Java and Python. All bugs are reproducible within the packaged container.

Defexts is a bug benchmark dataset aiming to collect bugs for the less popular JVM programming languages, namely Kotlin and Groovy \cite{Defexts-icse19}. The bugs are collected through finding a commit message with keywords related to bugs.

\citeauthor{refactory-ase19} collected 1783 bugs from 5 Python programming assignments in a Python introductory course \cite{refactory-ase19}. This dataset is then used to evaluate their tool for generating patches for student programs.

BugsInPy is a Python bug dataset, aiming to create a bug dataset similar to Defects4J, but for Python. 493 bugs are collected manually identify bug-fix commits, reproduce bugs with failing test cases and isolate the bug-fix change from other unrelated changes.

ManySStuBs4J is a collection of 153652 single statement bugs from 1000 popular Java projects \cite{manystupidbugs}. The bugs are curated by classifying bug-fix commits using keywords and filtering our clear refactors. The remaining bugs are then classified into 16 bug patterns.

\citeauthor{codit-tse20} gathered 32473 patches to train and evaluate their tree-based neural model for code editing \cite{codit-tse20}. The dataset is collected from 48 Java projects that used TravisTorrent, have at least 50 commits, and 10 watchers on GitHub.

\citeauthor{CoCoNuT} collected a multi-lingual bug dataset to train an ensemble model for program repair. The bugs are collected by keyword matching on commits. In total, they collected 23M samples for Java, Python, C, and JavaScript.

Megadiff is a collection of possible bug-fix commits for Java. Megadiff is curated by keyword matching on commit messages and filter commits that do not change Java or changed more than 40 lines of code.

Vul4J is a dataset of reproducible Java vulnerabilities \cite{vul4j2022}. All vulnerabilities from filtered from the Project KB knowledge base. The filter criteria include that it is Java, contains Java test suite, and is reproducible and isolated.

FixJS is a dataset of bug-fix commits for the JavaScript language \cite{fixjs-msr22}. The dataset is constructed by matching keywords on GitHub commits and filtering commits that did not change JavaScript files.

\approach stands apart from all the previously mentioned dataset construction approaches due to the following distinctions. Firstly, while the aforementioned datasets focus solely on codebases, \approach takes a different approach by connecting codebases to external bug repositories, allowing it to track precise bug information.
Secondly, \approach includes project-specific bugs generated by bug injector, which is not considered by any aforementioned datasets.

\subsection{Bug Injection}

Bug injection is most commonly employed in mutation testing to assess if an existing test suite captures a small source code change and to evaluate source code analyzing tools. Additionally, it has been utilized to generate training data for machine learning models focused on source code analysis. In this section, we will explore related works that involve creating artificial bugs.

Mutation testing has a lengthy history of altering source code to introduce bugs and verifying if the existing test suite detects them. For further information on mutation testing, we direct readers to two surveys \cite{jia2010analysis, papadakis2019mutation} as there are numerous studies on this subject. The main distinction between mutation testing and \approach bug injection lies in their focus. While mutation testing aims to identify bugs that do not cause test failures, \approach bug injection is employed to construct a bug dataset.

The Juliet test suite comprises over 81,000 synthetic C/C++ and Java programs with 181 categories of inserted faults \cite{boland2012juliet}. Developed by the National Security Agency’s Center for Assured Software (CAS), this test suite aims to evaluate the effectiveness of software assurance tools. However, the process by which the Juliet test suite creates these synthetic bugs remains unclear.

\citeauthor{shiraishi2015test} created 638 C/C++ programs with intentionally injected faults \cite{shiraishi2015test}. These injected faults are categorized into 9 defect types. The dataset is used to evaluate static analysis tools. However, the method employed by \citeauthor{shiraishi2015test} to create these bugs remains unclear.

LAVA is an automated system capable of injecting faults into large open-source C programs \cite{dolan2016lava}. The bugs are triggered by pre-defined inputs, whereas normal inputs are unlikely to trigger them. This system is used to evaluate bug-finding tools. LAVA differs from \approach in that it uses execution traces to insert bugs and only incorporates out-of-bound read/write bugs. On the other hand, \approach utilizes more rewrite results to create a wider variety of bug types.

EvilCoder is another automated tool that injects vulnerable code \cite{pewny2016evilcoder}. It identifies potentially vulnerable code locations and modifies them to become actually vulnerable by utilizing data flow analysis. This tool facilitates the systematic evaluation of bug-finding tools. EvilCoder varies from \approach in that the bugs it inserts are characterized by improperly secured data flow. In contrast, \approach employs more rewrite results to create a broader range of bug types.

Apocalypse is an automated bug injection tool based on symbolic execution \cite{roy2018bug}. It aims to inject realistic bugs, distinguishing itself from LAVA and EvilCoder. Apocalypse demonstrated its ability to generate diverse, difficult, and highly realistic bugs according to various metrics. The key difference between \approach and Apocalypse lies in the bug creation process. Apocalypse utilizes symbolic execution, whereas \approach relies on rewrite rules.

SemSeed introduced a novel method for automatically inserting realistic bugs \cite{patra2021semantic}. It employs machine learning techniques to learn patterns and characteristics from collected bug-fixing commits. The learned features are then used to seed new bugs into existing programs. The main distinction between \approach and SemSeed is in their bug creation approach. SemSeed relies on machine learning, while \approach uses rewrite rules.

BugLab is an approach that utilizes self-supervised training for bug detection and repairs \cite{allamanis2021self-buglab}. They create rewrite rules to artificially insert bugs into programs, which serve as training data. The rewrite rules used by BugLab include variable swap, argument swap, operator swap, and literal swap. The key difference between \approach and BugLab is the number of rewrite rules used. BugLab does not execute the injected bugs, therefore, no type and error message are obtained in their approach.

DrRepair is a graph-based program repair approach that also uses self-supervised training to generate artificial training data \cite{Yasunaga20DrRepair}. It randomly deletes, inserts, or replaces operators, punctuation, identifiers, and keywords to create bugs and record the compiler message. DrRepair is different from \approach in that it focus on compiler errors instead of bug that are exposed by the test suite.

SelfAPR is another example of using a bug injection model to create training data for machine learning models on program repair \cite{selfAPR2022}. They use 16 rewrite rules to create artificial bugs for training data, ensuring that all bugs are validated against the compiler and test suite.
\approach's rewrite rules are essentially derived from SelfAPR,
however, we  have re-formatted  them to construct a bug dataset that includes important bug location and error message information. On the contrary,  SelfAPR largely ignores these aspects as its goal is to  focus on  code change pattern learning.

\section{Conclusion}
We introduce a comprehensive and extensive bug-fix collection approach named \textit{\approach}, which encompasses three distinct sources for bug acquisition: CVEs from NVD, bugs from OSS-Fuzz, and injection-based bugs. This endeavor has resulted in a total of \numprint{1057818} bugs across \numprint{2968} open source projects. Notably, this dataset stands out as the largest bug-fix collection to date, encompassing precise bug types and accompanying message information, making it valuable for future software maintenance tasks, such as bug detection, fault localization, and automated program repair.

\approach offers solutions to create two types of bug-fix datasets: a general-wise dataset composed of real-world bug fixes made by developers, and a project-specific dataset that incorporates domain knowledge and aligns with the code style of the project. We believe that addressing the industry challenge of imprecise bug-fix datasets requires both components to build deep learning models that can learn broadly and in-depth.
In industry settings, where private and sensitive projects exist, having a project-specific bug-fix dataset becomes essential to enable training and learning from the same codebase while ensuring data security and privacy.
Furthermore, the flexibility of both components is significant as they are extensible, allowing for future expansion and adaptation.




\section{Acknowledgements}
We thank the anonymous reviewers for the insightful feedback.
This work was partially supported by The Wallenberg Foundation and WASP Postdoctoral Scholarship Program - KAW 2022.0368, and partially supported by the TrustFull project financed by the Swedish Foundation for Strategic Research. 
The computations and data handling were enabled by the supercomputing resource Berzelius provided by National Supercomputer Centre at Linköping University and the Knut and Alice Wallenberg foundation:Berzelius-2023-175.

\balance

\bibliographystyle{IEEEtranN}
\footnotesize
\bibliography{reference}

\end{document}